 \documentclass[final,5p,times,twocolumn,authoryear]{elsarticle}

\usepackage{amssymb}
\usepackage{lipsum}

\journal{}
\usepackage{amsmath}

\makeatletter
\long\def\@@address[#1]#2{\g@addto@macro\elsaddress{%
    \def\baselinestretch{1}%
     \refstepcounter{affn}
     \xdef\@currentlabel{\theaffn}
     \elsLabel{#1}%
    \textsuperscript{\theaffn}#2\par}}
\def\ps@pprintTitle{%
  \def\@oddhead{\reset@font{\hspace{5.6cm}[Revised copy, originally published in April 2020]}\hfil}
  \let\@evenhead\@oddhead
  \def\@oddfoot{\reset@font\hfil\thepage\hfil}
  \let\@evenfoot\@oddfoot
}
\makeatother

\begin{document}

\begin{frontmatter}



\title{Incentivized Third Party Collateralization for Stablecoins}


\author {Souradeep Das}

\author {Dr. Revathi Venkataraman}

\affiliation{organization={Computer Science and Engineering}, SRM Institute of Science and Technology}

\begin{abstract}
Stablecoins, which are primarily intended to function as a global reserve of value are insubstantial in their design and present many failure points. The primary mechanism to enable these coins to hold on to a fixed value is by backing them with collateral. Fiat collateralized stablecoins require users to trust a centralized entity, which breaks the total concept of decentralization. Crypto collateralized stablecoins have issues involving high collateral requirements and introduces risks of auto-liquidation. In this paper we aim to propose an alternative architecture for the creation of a functional and secure stablecoin.
\end{abstract}




\end{frontmatter}




\section{Introduction}
\label{introduction}

Stablecoins are cryptocurrencies designed to eliminate volatility by backing them with an asset or a currency that remains stable. To further specify - A stablecoin maintains its value in accordance with a fiat government backed currency. These special purpose tokens have garnered a lot of interest and appreciation due to their property of non-volatility, an attribute that has been a pre-dominant problem in cryptocurrencies of today. 
     However, Stablecoin designs which exist currently are insubstantial and present several failure points, limiting their general acceptability. Fiat collateralized stablecoins require users to trust a central entity, which breaks the entire concept of decentralization and self-reliance. Non-collateralized stablecoins implementing seigniorage share approaches are still new and users putting their trust on them is questionable. Slightly better off - the Crypto-Collateralized stablecoins, although providing a decentralized ecosystem are surprisingly not very efficient in maintaining stability and moreover, have issues in the form of high collateral requirements and liquidation risks during market downturns. For instance, Crypto-backed stablecoins like Dai from Maker, have achieved the feat of decentralization in the system [1], but still contain limitations in their design principles in the form of–

    \begin{itemize}
        \item requirement of ~1.5 x collateral
        \item Collateral can be auto liquidated
    \end{itemize}

All these aspects give rise to a dire need and opportunity to improve this piece of technology for global reach and acceptance.

\section{State of Existing Problems}

\subsection{Artificial Markets}

The market is dependent on supply-demand relationships. As several underlying cryptocurrencies supporting stablecoins are native coins in their own chain, proof of work is the way to mine and produce new coins for the market. While the generation of such new coins is adjusted to happen at regular intervals, and in turn, the mining difficulty of such blockchains is adjusted, this still produces an artificial market to base a value of a stablecoin. This calls for a separation of concerns between a stablecoin and the tokenomics of its underlying cryptocurrency. [2].

\subsection{Market Manipulation}
Market manipulation techniques allow any entity to try and manipulate or alter the price of the stablecoins. In stark contrast to the effects of manipulating general cryptocurrency markets, controlling or influencing stablecoins could be a cause of destroying the functionality of an entire ecosystem. This could result in an improper game to play out and could take down the underlying value and market instantly. Furthermore, the existential risk of market collapse also destroys the whole concept of a decentralized cryptocurrency.

\section{An Update to the Architecture}
These existing problems all direct to a new form of support the architecture of stablecoins require. One way of preserving all the benefits while solving the problems is by creating  an additional collateral backing from third-party investments. Broadly speaking, a crypto-collateralized stablecoin where the collateral backing is from third party investments would allow a 1:1 collateralization ratio, while also enabling the use of the surplus investor funds as a security if/when the price of Ethereum goes down. The stability, operation and incentives of the system is to be taken care of by a competitive investment process.

\subsection{The process}

Users deposit (ETH) and receive an equivalent amount of stablecoins in return of the collateral provided. However, unlike DAI and other crypto backed stablecoins, no extra amount of ether over the stablecoin amount has to be provided by the user. After the exchange, the stablecoin amount will need to be secured against volatility in the underlying crypto (Ethereum) by an additional group of people. This external crowdfunded pool ensures a transfer of  the liability of the volatile collateral asset to the pool funders instead of the stablecoin buying user. The members of this pool, in return of securing the stablecoin are entitled for the profit (or loss, if any) in denominations of the underlying crypto and this helps keep the stablecoins stable.

\subsection{Incentives for Investment}

The algorithm proposed later, provides better returns to the investors for providing a portion of the collateral on the external pool, than just holding on to their Ethereum (ETH)[3]. The investment made by the investor is returned (along with rewards) when the original stablecoins are redeemed. The amount returned to each of the investors is not linear to the fraction of collateral they provide, but exponential depending on their standing. The algorithm further ensures incentive generation on the external pool is self-sufficient in its distribution. While the external pool makes up for any missing user collateralization, the funds pooled in on the external pool are the same funds redistributed to the investors in ratios determined by the algorithm.

\subsection{External Collateral provision from the Pool}

Providing the collateral for the stablecoin through a method of external funding backed by incentives offers for a structure which has two independent parts that function together to make the system functional. The external collateral pool being a separate entity for controlling the stability ensures that the instability of the market is not a concern for the proper functioning of the application. Separating the collateral pool and disconnecting all relations with the other aspects of the stablecoin structure also allow the system to adapt to an alternate method of gathering collateral, if required. This increases the security aspect and makes sure that the system has a 100\% uptime.

\section{Explaining competition}

\subsection{Anonymity to enforce competition}

The portions filled by other investors are anonymous. This provides a competitive glance to filling up the maximal portion of the collateral amount. After all the investors are done filling up the collateral, an internal threshold is formed which partitions the point of profit/loss on either side.

\subsection{Risk/Reward Gains}
The anonymity brings in a concept of a strategic game built into the system. The collateral providers are provided to maximize their lending capacity considering the risk factor to it. Since the rewards are directly related to the amount of investments, without the knowledge of other investor’s pooled in amounts, each investor would optimize to try and provide the maximal amount into the pool. The only risk factor is the criteria for slashing the funds if a volatility issue arises.    

\begin{figure}
	\centering 
	\includegraphics[width=0.48 \textwidth]{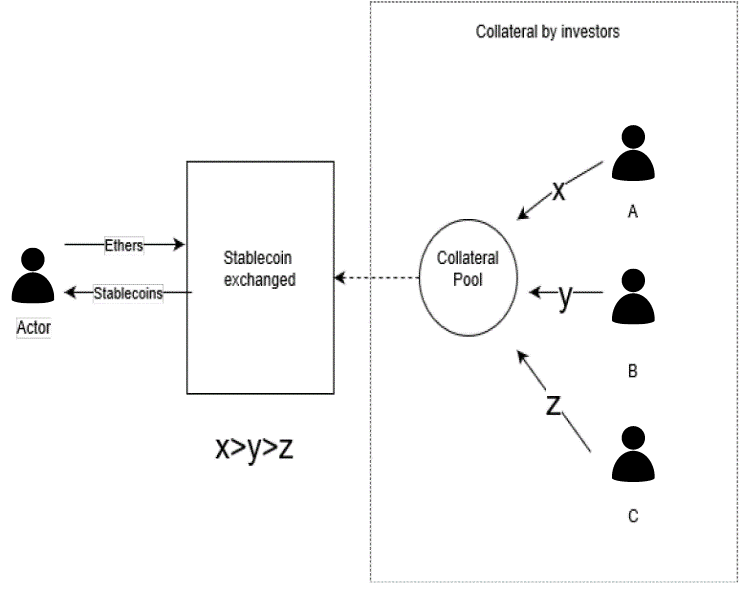}	
	\caption{Collateral Pool structure} 
	\label{fig_pool}%
\end{figure}

\subsection{Keeping the system alive}

The competition among the lenders ensures a faster disbursal rate and provides an almost instantaneous rate for providing with the capital/collateral. While crypto-collateralized stablecoins currently in existence provide a method to collateralize instantly, this mechanism comes close to that when the scale gets bigger. This forms a three ended cycle where the speed of disbursal increases with competition, which is further increased by trust in the system.

\section{Explaining the Money-Money Algorithm}

\subsection{Designing the algorithm}

The Algorithm takes care of the return distributions to the investors of the collateral pool. A strong and intelligent mechanism would motivate the investors and borrowers (the token buyers) to play a strategic game together and keep the system functional. The algorithm tends to provide returns in an exponential form with respect to the amount invested or the rate of involvement in the system.

\subsection{Description of the design}


The Algorithm shows the Distribution of the 
MARGIN (CURRENT ETH VALUE (-) STABLECOIN PURCHASED)
in an Exponential way proportional to the Amount invested
\\\\
Let,
\\[0.3\baselineskip]
Incentive[i] =  Incentive of individual investor
\\[0.3\baselineskip]
Lsum = Total amount of collected incentives
\\[0.3\baselineskip]
filled[i] = Portion filled by ith person
\\[0.3\baselineskip]
T = Total Limit
\\[0.3\baselineskip]
A = Amount cumulated
\\[0.3\baselineskip]
\\
The intermediary incentive of each individual -
\begin{equation}
\text{incentive}[i] = \frac{e^{(\text{filled}[i]-T)}}{\text{filled}[i]}
\end{equation}

The total amount of incentives:-

\begin{equation}
L_{\text{sum}} = \sum_{i=1}^{n} \text{incentives}[i]
\end{equation}

Calculating the individual gain fractions of each investor-
\begin{equation}
\text{finalIncentiveFractional}[i] = \frac{\text{incentive}[i]}{L_{\text{sum}}}
\end{equation}

The return from the collateral building:-
\begin{equation}
\text{finalIncentive}[i] = \frac{\text{incentive}[i]}{L_{\text{sum}}} \cdot A
\end{equation}

\subsection{Optimal Strategy}

The core of the algorithm is distributing pooled token amounts in relation to the parts filled by each investor. As each investor tries to fill maximal portions of the pool, the returns generated are always higher for all investors over the inflection point. Hence, the optimal strategy is to fill the highest fraction of the collateral pool amount to get the most significant profits (if any). The competition among the investors for landing on the better side of the curve (filling the most parts of the collateral), will also mean faster fulfillment for potential collateral balancing.

To further elaborate the exact advantages of the investors, we list out the two cases (as illustrated in Fig 2) -
    \begin{itemize}
         \item \textit{When ETH prices go up, people above the optimal point receive a higher profit than what they could have got by simply holding the ETH}

      \item \textit{When ETH prices go down, loss is significantly less compared to what they would have incurred by simply holding the ETH, given they are above the optimal point}

    \end{itemize}

\section{Statistics}

The stablecoins structure proposed is non-linear type when it comes to the returns offered against investments. Because of the exponential structure of the profits there exists a threshold which has to be crossed in order to attain the profits. The profits are the reason which forces the lenders to compete for the extra rewards, which in turn motivates everyone to increase their contributions to the pool and increase the stablecoin disbursal capacity. 

Sample crowdfunded lending data was taken, and results were compared to find incentives to use our proposed model over a flat interest distribution model.
The red graph (Fig 2) plots the investments to returns for holding on to a cryptocurrency like Ether (ETH), while the blue graph plots the performance of the proposed platform. The point of intersection (at the center) specifies the central point or the threshold to land with extra rewards. The results indicate - whether it’s a bull or bear crypto market, investing in the collateral pool will mean greater profits or lesser losses for investors, and strengthen a stablecoin architecture simultaneously.

\begin{figure}
	\centering 
	\includegraphics[width=0.48 \textwidth]{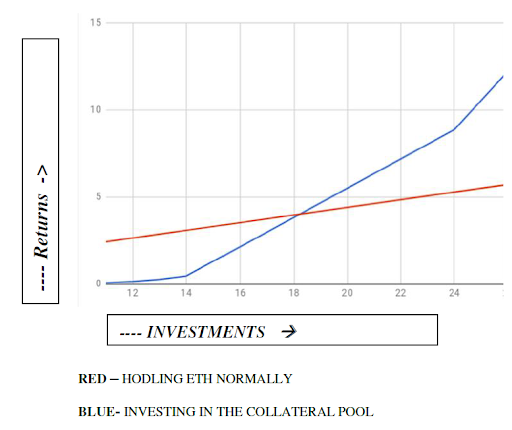}	
	\caption{Comparing Incentives vs Investments} 
	\label{fig_com}%
\end{figure}

\section{Additional Integrations}
\subsection{Introduction of a Secondary token}

A secondary token could be introduced to the ecosystem which could bind both the tokens together when maintaining stability. The secondary token could also be traded in a similar manner with the differences only being in the usage, to-
\begin{itemize}
    \item Adjust the supply and demand manually by investors to maintain the stability of the primary currency
    \item Be used as a token to prevent or distribute losses in a better way in the situation of a high risk volatility
\end{itemize}

The potential of two tokens working together in maintaining stability was first furnished through the inflation adjusting ‘maker’ token in the DAI ecosystem.

\subsection{Adding a lending service to integrate with the ecosystem}

The stablecoin can be considered to be a loan itself, since stablecoins are provided in exchange for a collateral. The ecosystem already uses a method to provide collateral by third-party or entities who are not acquiring the stablecoins. The function of the collateral can be expanded to incorporate a lending structure [4] to make up for the losses by the market volatility. The amount of existent tokens in the collateral pool can serve as assets that can be lent to make up for losses (during market downturns)  in the system. The utility of the collateral pool can also be shared with a microfinance entity, or a decentralized lending protocol . This however, could bring up some issues, namely-

    \begin{itemize}
        \item Introduce a separate entity of trust within the collateral pool
        \item Provide a point of liquidity risk when collateral pool is depleted enough, lower than the market capacity [5]
    \end{itemize}

\subsection{Integrating zero-knowledge privacy}

A core aspect of the collateral pool design is the anonymity of investor contributions. Anonymity enforces contributions when investors compete to fill the maximum portions of the collateral pool. Hiding the individual contribution can be achieved through a commit-reveal scheme, where each investor makes a commitment of their contribution, and reveals the amount only when funds by all investors have been pooled in. A better way to realize this could be through hiding commitments on-chain and integrating zero-knowledge (ZK) computations for proving the algorithm. Designing a strong model would further ensure the preserved privacy of investors and ensure a fair and competitive game.

\section{Implementation Discussion}
The primary components of the ecosystem include:-
\begin{itemize}
    \item Ethereum Smart Contracts
    \item ERC-20 token
    \item Wallet
    \item Interest distribution Algorithm
\end{itemize}

The stablecoin has been designed to be functional at the Ethereum main network as an ERC-20 token standard. The token was created using solidity smart contracts and deployed to the Ethereum network. A secondary token can be added to maintain or further improve the stability of the existing token. The secondary token is the backing token and could also be an investment medium for several third parties. Smart contracts are the piece of code that lives on the world-computer- the Ethereum Blockchain and are the way to implement these tokens on the Ethereum framework. The process of using the token i.e. buying and transferring can be done by interacting with the contract. Instead of direct interactions with the contract, users could also prefer using a wallet service to avoid directly using the function calls of the contract. The application has been tested by creating a wallet and complementary scripts that provide a medium to interact and perform the necessary functions.

\section{Conclusion}
This solution provides a reliable and unique strategy to encourage more people into securing a stablecoin architecture and maintain its peg to the value of a fiat currency. This is enforced by the competition in pooling assets as the collateral, and incentives in the form of meaningful returns from the system. This not only helps in creating a stable ecosystem and currency structure but also increases the economic activity and currency flow [6]. 

Additionally, the platform also ensures total functionality in all the discovered scenarios with proper integrity and completeness throughout all the components in this proposed system. The proposed system can hence be easily integrated to an existing stablecoin structure or be utilized afresh.


\end{document}